\documentclass{PoS}

\usepackage{amssymb,amsmath,wrapfig}
\title{Supersymmetry breaking on the lattice:\\ the ${\cal N}=1$ Wess-Zumino model}
\ShortTitle{The ${\cal N}=1$ Wess-Zumino model on the lattice}

\author{David Baumgartner, Kyle Steinhauer and \speaker{Urs Wenger} \\
        Albert Einstein Center for Fundamental Physics\\
         Institute for Theoretical Physics\\
        University of Bern\\
        Sidlerstrasse 5\\
        CH--3012 Bern\\
        Switzerland\\
        E-mail: \email{baumgart@itp.unibe.ch}, \email{steinhauer@itp.unibe.ch},
        \email{wenger@itp.unibe.ch}}

      \abstract{We discuss spontaneous supersymmetry breaking in the
        ${\cal N}=1$ Wess-Zumino model in two dimensions on the
        lattice using Wilson fermions and the fermion loop
        formulation. In that formulation the fermion sign problem
        related to the vanishing of the Witten index can be
        circumvented and the model can be simulated very efficiently
        using the recently introduced open fermion string
        algorithm. We present first results for the supersymmetry
        breaking phase transition and sketch the preliminary
        determination of a renormalised critical coupling in
        the continuum limit.}

\FullConference{ The XXIX International Symposium on Lattice Field Theory - Lattice 2011\\
                 July 10-16, 2011\\
                 Squaw Valley, Lake Tahoe, California}

\newcommand{\dslash}{\partial \hspace{-6.25pt}\slash}

\newcommand{\be}{\begin{equation}}
\newcommand{\ee}{\end{equation}}

\newcommand{\beann}{\begin{eqnarray*}} 
\newcommand{\eeann}{\end{eqnarray*}}

\begin{document}

\section{Introduction}
Supersymmetry as an extension of the space-time symmetries is an
interesting concept which might well be realised in nature in one form
or another. Its presence has several intriguing consequences, for
example, one expects a vanishing ground state energy, if supersymmetry
is exact. Moreover, the particle spectrum contains mass degenerate
bosons and fermions which are related by the supersymmetry. However,
so far there has not been any experimental sign of such a
boson-fermion degeneracy. So, if supersymmetry is indeed realised at
some high energy scale, it must be broken at the low energy scales
accessible in today's experiments. The scenario of (spontaneously)
broken supersymmetry then implies that there is no supersymmetric
ground state, the ground state energy is not vanishing, and the
particle masses need not be degenerate.  In this context, it is
interesting to ask how the spontaneous breaking of the supersymmetry
is realised. Since spontaneous symmetry breaking is an inherently
non-perturbative problem one needs non-perturbative methods in order
to approach it meaningfully. One such method is provided by simulating
supersymmetric theories on a space-time lattice. However, since the
space-time symmetries are explicitly broken by the lattice
regularisation and are restored only in the continuum limit, also the
supersymmetry is in general not (or not fully) realised on the
lattice.  As a consequence, there is a subtle and delicate interplay
between the various symmetries, and their realisation in the continuum
needs to be carefully studied.
Here we present preliminary results of such a study for the ${\cal
  N}=1$ Wess-Zumino model in two dimensions. Using the Wilson fermion
discretisation one can formulate the model in terms of fermion loops
which can be simulated very efficiently using the open fermion string
algorithm \cite{Wenger:2008tq}. In addition, the fermion loop
formulation provides a way to circumvent the sign problem related to
the vanishing of the Witten index \cite{Baumgartner:2011cm}.

\section{The  ${\cal  N}=1$ Wess-Zumino model}
The ${\cal N}=1$ Wess-Zumino model in two dimensions is one of the
simplest models which may exhibit spontaneous supersymmetry breaking.
Its degrees of freedom consist of one real Majorana fermion field
$\psi$ and one real bosonic field $\phi$, while its dynamics is
described by the Lagrangian density
\be
\label{eq:WZLagrangian}
{\cal L} = \frac{1}{2} \left(\partial_\mu \phi \right)^2
+ \frac{1}{2} P'(\phi)^2 
+ \frac{1}{2}\overline{\psi} \left( \dslash + P''(\phi) \right) \psi
\, .
\ee
Here, $P(\phi)$ denotes a generic superpotential, and $P', P''$ its
first and second derivative with respect to $\phi$. In the following
we will concentrate on the specific form  
\be
\label{eq:cubic superpotential}
P(\phi) = \frac{m^2}{4 g} \, \phi + \frac{1}{3} g \phi^3
\ee
which leads to a vanishing Witten index $W=0$ and hence allows for
spontaneous supersymmetry breaking \cite{Witten:1982df}.  The
corresponding action enjoys the following two symmetries. First, there
is a single supersymmetry given by the transformations
\be
\delta \phi = \overline{\epsilon} \psi \, , \quad 
\delta \psi = (\dslash \phi - P') \epsilon \, , \quad
\delta \overline{\psi} = 0 \, , 
\ee
and second, there is a discrete  $\mathbb{Z}(2)$ chiral symmetry given
by
\be
\phi \rightarrow  -\phi \, , \quad
\psi \rightarrow \gamma_5 \psi  \, , \quad 
\overline{\psi} \rightarrow  - \overline{\psi} \gamma_5 \, ,
\ee
where $\gamma_5 \equiv \sigma_3$ can be chosen to be the third Pauli
matrix.

For the chosen superpotential the Witten index turns out to be zero,
as can be seen as follows.  Integrating out the Majorana fermions
yields the (indefinite) Pfaffian $\text{Pf} \, M$ of the Majorana
Dirac operator $M$. The corresponding partition function with periodic
boundary conditions (b.c.) in all directions is equivalent to the
Witten index,
\[
\int {\cal D}\phi \, e^{-S_\text{b}(\phi)}\,  \text{Pf} \, M_\text{pp}(\phi)
\propto W \,,
\]
where $S_\text{b}(\phi)$ is the action for the bosonic field.
Now, under the $\mathbb{Z}(2)$ symmetry $\phi \rightarrow  -\phi$ one
has
\[
S_\text{b} \rightarrow S_\text{b}, \quad \text{Pf}\, M_\text{pp} \rightarrow -
\text{Pf}\, M_\text{pp} \, ,
\]
so for every bosonic field configuration $\phi$ contributing to the
partition function, there exists another one with exactly the same
contribution but opposite sign, hence yielding $W=0$. This constitutes
a necessary (but not sufficient) condition for the supersymmetry to be
broken spontaneously. In that case, one expects a bosonic and
fermionic ground state related to each other by the supersymmetry
transformation.
On the other hand, if one chooses thermal b.c.~(antiperiodic b.c.~for
the fermions in time direction) the supersymmetry is broken by the
finite temperature of the system and one finds
\[
S_\text{b} \rightarrow S_\text{b}, \quad \text{Pf}\, M_\text{pa} \rightarrow +
\text{Pf}\, M_\text{pa} \, .
\]

In order to further understand the supersymmetry breaking pattern,
i.e.~the relation between the supersymmetry breaking and the
$\mathbb{Z}(2)$ symmetry breaking, it is useful to consider the
potential for the bosonic field, $\frac{1}{2} P'(\phi)^2 = \frac{1}{2} \frac{m^2}{2} \phi^2 +
\frac{1}{2} g^2 \phi^4 + \text{const}$.
It simply represents a standard $\phi^4$-theory in which, depending on
the choice of the bare parameters $m$ and $g$, the $\mathbb{Z}(2)$
symmetry may be broken. Indeed, for large values of $m/g$ the
$\mathbb{Z}(2)$ symmetry is spontaneously broken (in infinite volume)
and the boson field selects a definite ground state. Denoting with
$\overline \phi$ the expectation value of the volume averaged boson
field, one finds
\beann
\overline \phi = + m/2g & \quad \Rightarrow \quad& \text{Pf}\, M_\text{pp} = +\text{Pf}\, M_\text{pa}
\, , \\
\overline \phi = - m/2g & \quad \Rightarrow \quad& \text{Pf}\, M_\text{pp} = -\text{Pf}\, M_\text{pa}
\, , 
\eeann
i.e.~in the former case the unique ground state is bosonic, while in
the latter it is fermionic. In both cases, there is a single, unique
ground state tantamount to having unbroken supersymmetry.
By contrast, for small values of $m/g$ the $\mathbb{Z}(2)$
symmetry is unbroken, i.e.~one has $\overline \phi = 0$ which allows
both a bosonic and fermionic ground state, tantamount to having broken
supersymmetry. Indeed, the tunneling between the two equivalent ground
states corresponds to the massless Goldstino mode which comes along
with any spontaneous supersymmetry breaking.

\subsection{Lattice discretisation and fermion loop formulation}
In order to put the Wess-Zumino model on the lattice we follow the
approach of Golterman and Petcher \cite{Golterman:1988ta} where it is
shown that using the same lattice derivative for the bosons as for the
fermions (and renormalising the mass parameter $m$ accordingly), the
supersymmetry is guaranteed to be restored in the continuum limit.

Using the Wilson lattice discretisation for the fermion fields yields
the fermion Lagrangian density
\[
 {\cal L} = \frac{1}{2} \xi^T {\cal C} (\gamma_\mu \tilde \partial_\mu
 - \frac{1}{2} \partial^* \partial + P''(\phi)) \xi \, ,
\]
where $\xi$ is a real, 2-component Grassmann field, ${\cal C} = -{\cal
  C}^T$ is the charge conjugation matrix and $\partial^*, \partial$
are the backward and forward lattice derivatives,
respectively. However, while the Wilson term $\partial^* \partial$
avoids fermion doubling, it spoils the discrete chiral symmetry of the
fermion action as well as the $\mathbb{Z}(2)$ symmetry $\phi
\rightarrow -\phi$ of the boson action\footnote{In principle this
complication can be avoided by using a lattice discretisation which
respects the discrete $\mathbb{Z}(2)$ chiral symmetry, e.g.~the SLAC
derivative \cite{Wozar:2011gu}.}.

Another problem for simulating the model on the lattice is
the fact that the Pfaffian is indefinite. As discussed above, this is
due to the vanishing of the Witten index and constitutes a generic
problem for any numerical Monte Carlo investigation of spontaneous
supersymmetry breaking, independent of the chosen discretisation.  The
problem stems from the fact that the indefinite Paffian can not be
simulated directly by standard Monte Carlo methods. Instead one uses
the effective action 
%
\[
S_\text{eff}(\phi) = S_\text{b}(\phi) - \ln |\text{Pf} \, M(\phi)|
\]
for the boson field $\phi$ and takes the sign of the Pfaffian into
account by reweighting. This approach in general leads to severe sign
problems \cite{Wozar:2011gu,Catterall:2003ae}.

It turns out that the sign problem can be circumvented for Wilson
fermions by using an exact reformulation of the lattice model in terms
of fermion loops \cite{Baumgartner:2011cm} and simulating fluctuating
fermionic boundary conditions \cite{Wenger:2008tq}. In the loop
formulation one expands the Boltzmann factor of the fermion action,
effectively constructing a hopping expansion. When one subsequently performs
the integration over the fermion fields, the nil-potency of the
Grassmann elements ensures that only closed, non-oriented and
non-intersecting fermion loops survive. The partition function then
becomes a sum over all self-avoiding fermion loop configurations
$\ell$,
\[
Z_{\cal L} =  \sum_{\{\ell\}\in{\cal L}} \omega[\ell,\phi] ,\quad
{\cal L} \in {\cal L}_{00} \cup {\cal L}_{10} \cup {\cal L}_{01} \cup {\cal L}_{11}
\]
where $\omega[\ell,\phi]$ denotes the weight for a given loop
configuration $\ell$, and ${\cal L}_{ij}$ denotes the equivalence class
of loop configurations with an even or odd number of loops winding
around the lattice in the spatial and temporal direction,
respectively. $Z_{\cal L}$ represents a system with
unspecified fermionic b.c. \cite{Wolff:2007ip}, while the system with
periodic b.c.~for the fermion, i.e.~the Witten index, can be
constructed by forming
\[
W \equiv Z_\text{pp} = Z_{{\cal L}_{00}} - Z_{{\cal L}_{10}} - Z_{{\cal L}_{01}} -
Z_{{\cal L}_{11}} \, ,
\]
or the system at finite temperature by forming
\[
\phantom{W \equiv} Z_\text{pa} = Z_{{\cal L}_{00}} - Z_{{\cal L}_{10}} + Z_{{\cal L}_{01}} +
Z_{{\cal L}_{11}} \, .
\]
Note that the weight $\omega$ does not necessarily need to be positive
definite in each of the sectors, but in practice it turns out that it
is the case as long as one stays close enough to the continuum limit.

As described in \cite{Wenger:2008tq} the system can most efficiently
be simulated by introducing an open fermion string corresponding to
the insertion of a Majorana fermion pair. By letting the ends of the
string move around the lattice by a standard Metropolis update
procedure, one samples the fermion 2-point function as well as the
relative weights between $Z_{{\cal L}_{00}}, Z_{{\cal L}_{10}},
Z_{{\cal L}_{01}}$ and $Z_{{\cal L}_{11}}$. Finally, the bosonic
fields are simulated by standard Monte Carlo methods.

\section{Results}
\begin{figure}[t]
\includegraphics[width=0.5\textwidth]{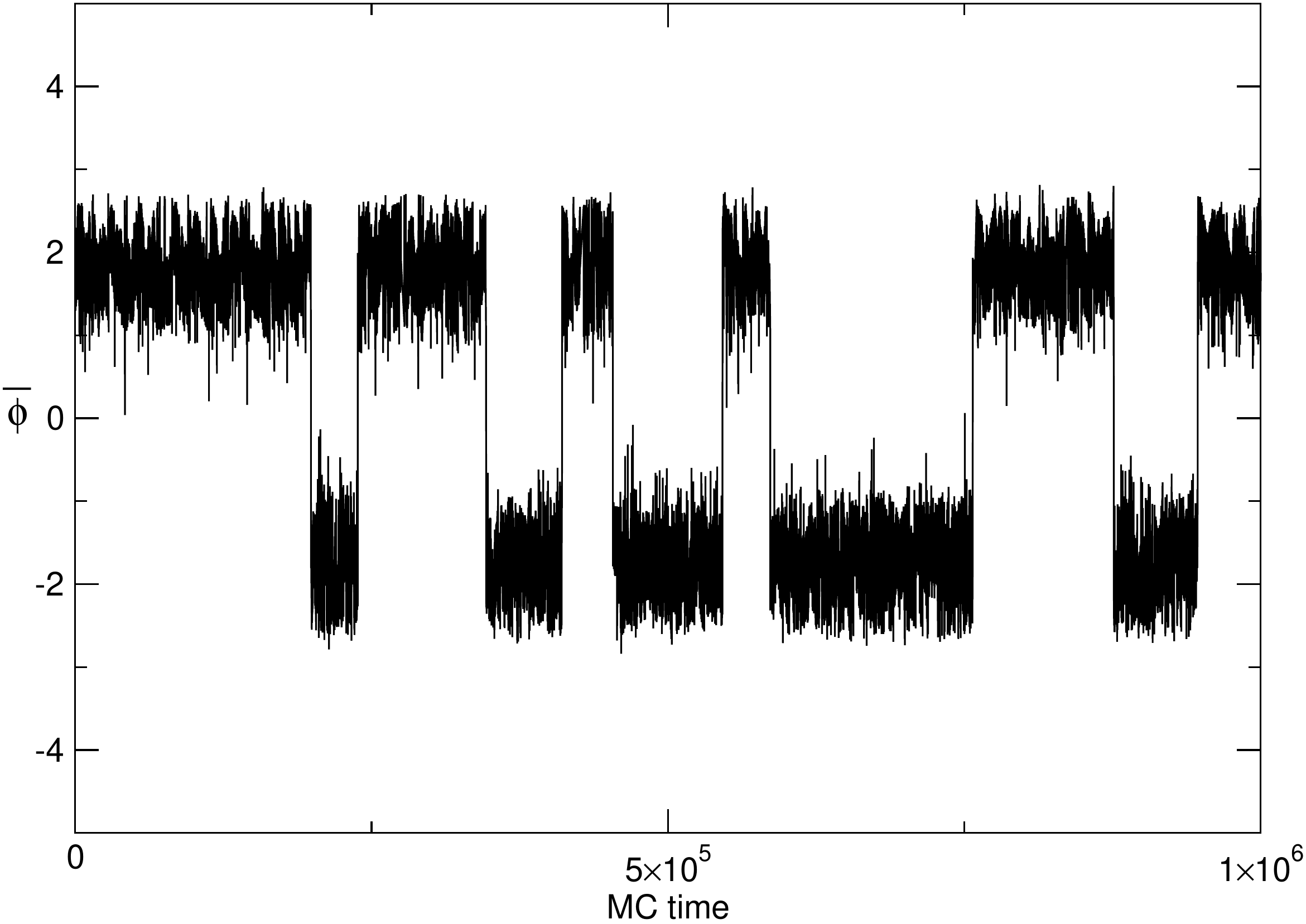} 
\includegraphics[width=0.5\textwidth]{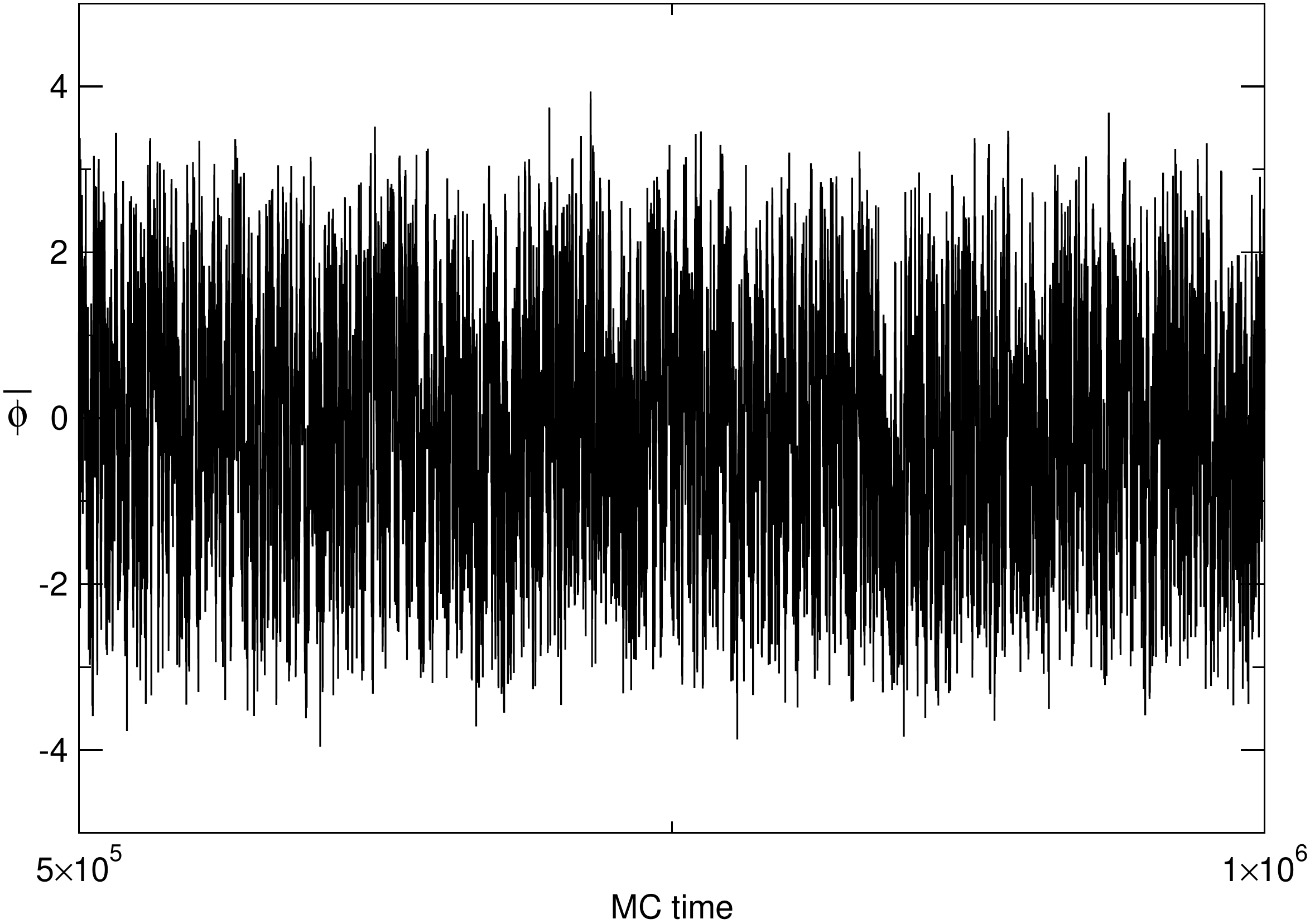} 
\caption{Monte Carlo time history of $\overline \phi$ at $m/g = 4,\, a
  g = 0.125$ (left plot) and $m/g = 0.16,\, a g =
  0.03125$ (right plot). \vspace{-0.3cm} }
\label{fig:MCtimehistory}
\end{figure}

By looking at the behaviour of $\overline \phi$ for large and small
values of $m/g$ one can check for the $\mathbb{Z}(2)$ symmetry breaking\footnote{Note that
  $\overline \phi$ is not a true order parameter for the
  $\mathbb{Z}(2)$ symmetry since the symmetry is explicitly broken by
  the lattice discretisation.}.
The plots in
Fig.~\ref{fig:MCtimehistory} show the Monte Carlo time history of
$\overline \phi$ at $m/g = 4, \, a g = 0.125$ (left plot) and $m/g =
0.16, \, a g = 0.03125$ (right plot). For $m/g = 4$ the system is in
the $\mathbb{Z}(2)$ broken phase with $\langle |\overline \phi|
\rangle \simeq 2$. Since the system is at finite volume it still
tunnels between the two vacua with $\langle |\overline \phi| \rangle \simeq
\pm m/2g$, but the tunnelling will be suppressed in the limit $L\rightarrow
\infty$ or $m/g \rightarrow \infty$. For $m/g = 0.16$ on the other
hand, the system is in the $\mathbb{Z}(2)$ symmetric phase with
$\langle \overline \phi \rangle \simeq 0$.

\begin{figure}[h]
\includegraphics[width=0.5\textwidth]{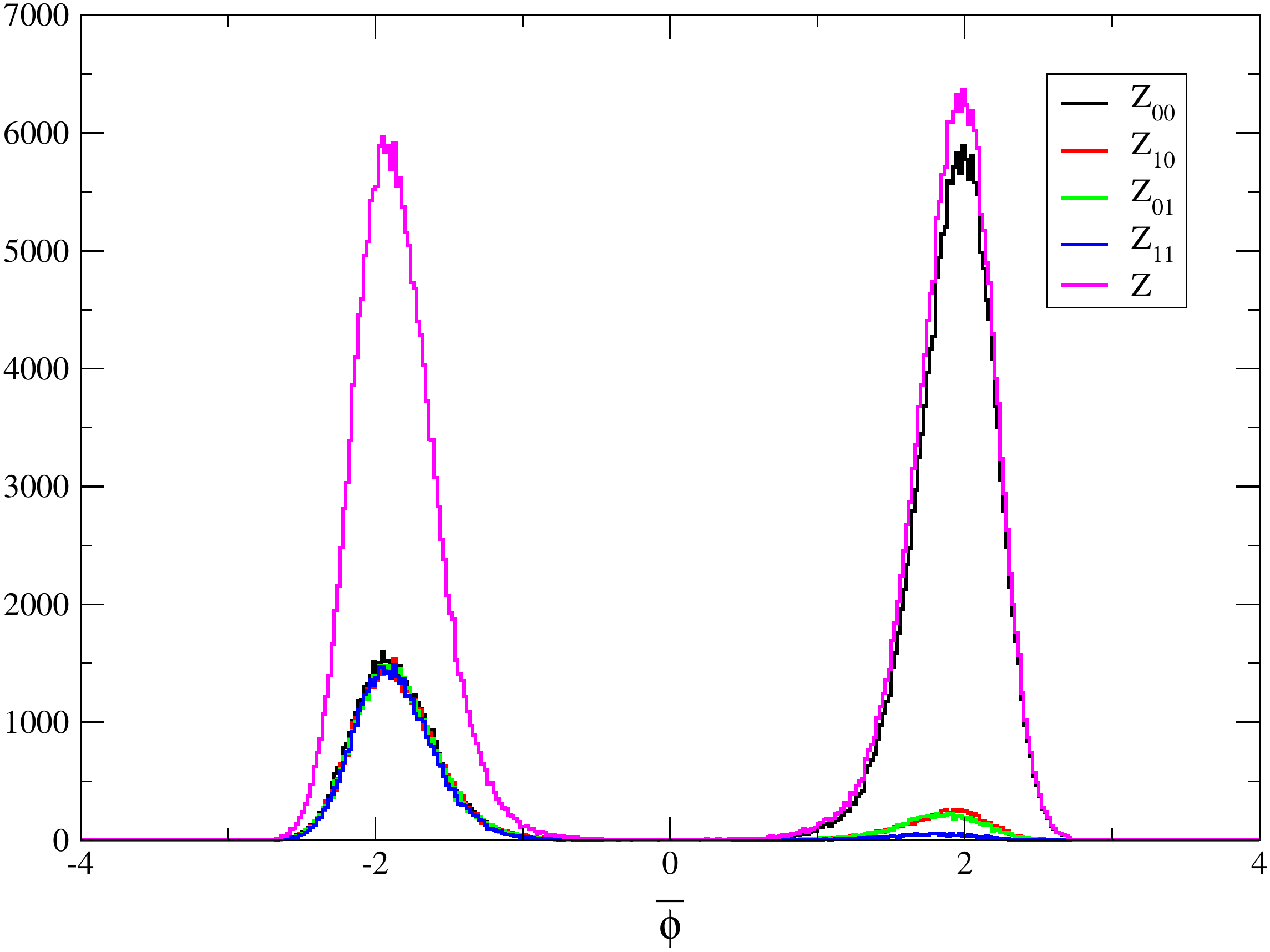} 
\includegraphics[width=0.5\textwidth]{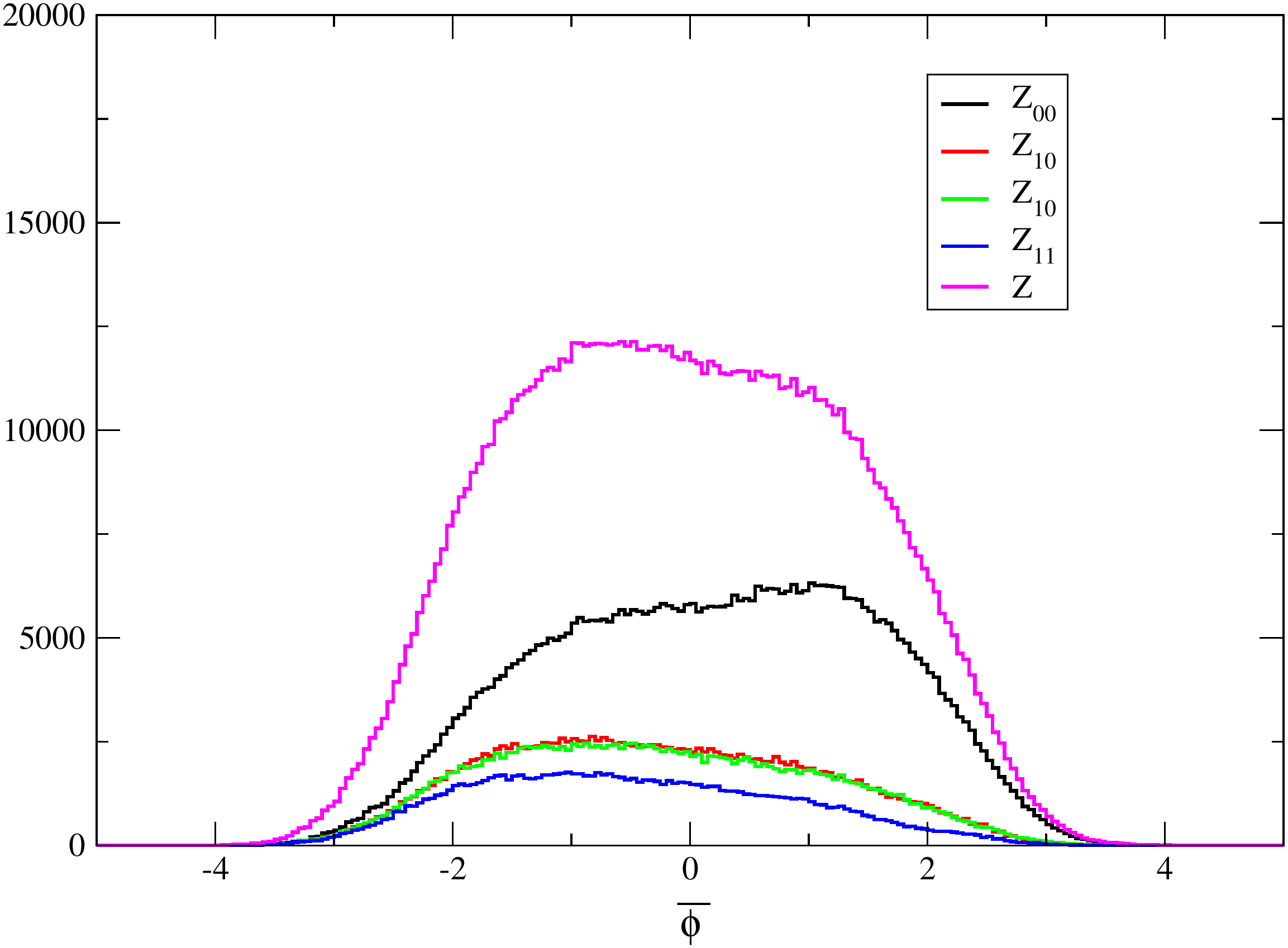} 
\caption{Probability distribution of the partition functions $Z_{{\cal
      L}_{00}}, Z_{{\cal L}_{10}}, Z_{{\cal L}_{01}}, Z_{{\cal
      L}_{11}}$ and $Z_{\cal L}$ for $m/g = 4, \, a g = 0.125$ (left
  plot) and $m/g = 0.16, \, a g = 0.03125$ (right plot) as a function
  of $\overline \phi$.\vspace{-0.3cm}}
\label{fig:Zdistributions}
\end{figure}

It is now interesting to see how the partition functions $Z_{{\cal
    L}_{00}}, Z_{{\cal L}_{10}}, Z_{{\cal L}_{01}}, Z_{{\cal
    L}_{11}}$, or $Z_\text{pp}$ and $Z_\text{pa}$, behave in the two
situations. Fig.~\ref{fig:Zdistributions} shows the probability
distributions of the various sectors as a function of $\overline \phi$,
again for $m/g = 4, \, a g = 0.125$ (left plot) and $m/g = 0.16, \, a
g = 0.03125$ (right plot). In the first situation where the
$\mathbb{Z}(2)$ symmetry is broken, one finds
\[
\begin{array}{clcl}
\langle \overline \phi \rangle
\simeq -2: \quad& Z_{00} \simeq Z_{10} \simeq Z_{01}
\simeq Z_{11} & \quad \Rightarrow \quad & Z_\text{pp} \simeq
-Z_\text{pa} \, ,\\
\langle \overline \phi \rangle
\simeq +2: \quad & Z_{00} \simeq 1, \, \,  Z_{10} \simeq
Z_{01} \simeq Z_{11} \simeq 0 & \quad \Rightarrow \quad & Z_\text{pp}
\simeq +Z_\text{pa} \, ,
\end{array}
\]
so $\langle \overline \phi \rangle \simeq -2$ corresponds to the
fermionic ground state while $\langle \overline \phi \rangle \simeq
+2$ corresponds to the bosonic one. In either case, a unique ground
state is chosen by the system (up to finite volume tunneling) and
hence supersymmetry is unbroken (at least in the thermodynamic and
continuum limit).

In the second situation, where the $\mathbb{Z}(2)$-symmetry is unbroken, one finds
\[
\begin{array}{clcl}
\langle \overline \phi \rangle
\simeq 0: \quad& Z_{00} \simeq Z_{10} + Z_{01} +
Z_{11} & \quad \Rightarrow \quad & Z_\text{pp} \simeq
 0 \, ,
\end{array}
\]
so the bosonic and fermionic ground states occur with equal
probability (thereby cancelling their contribution in $Z_\text{pp} =
W$) and hence supersymmetry is spontaneously broken.

As a next step one can now determine the value of $m/g$ at which the
\begin{figure}[b]
\includegraphics[width=0.5\textwidth]{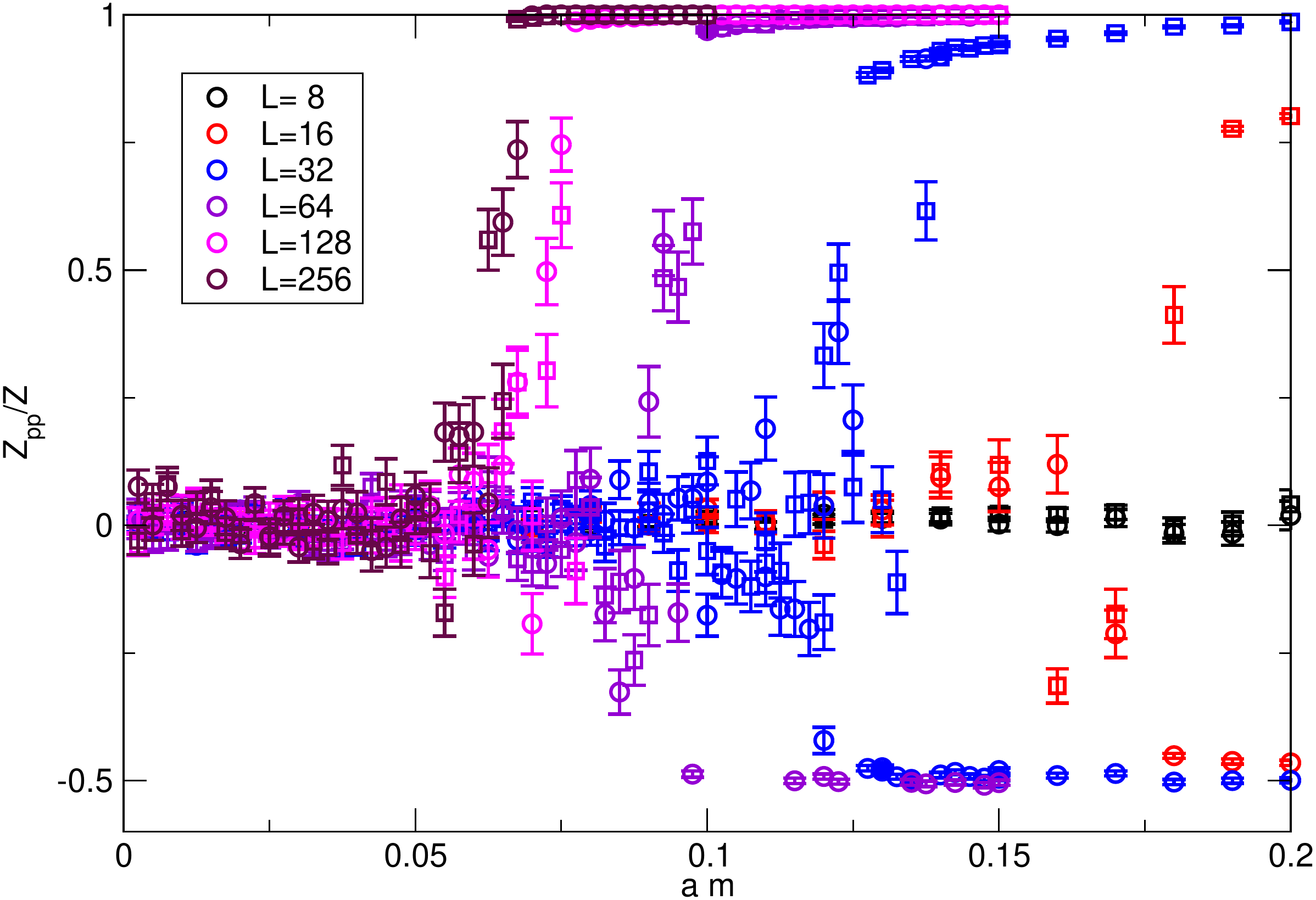} 
\includegraphics[width=0.5\textwidth]{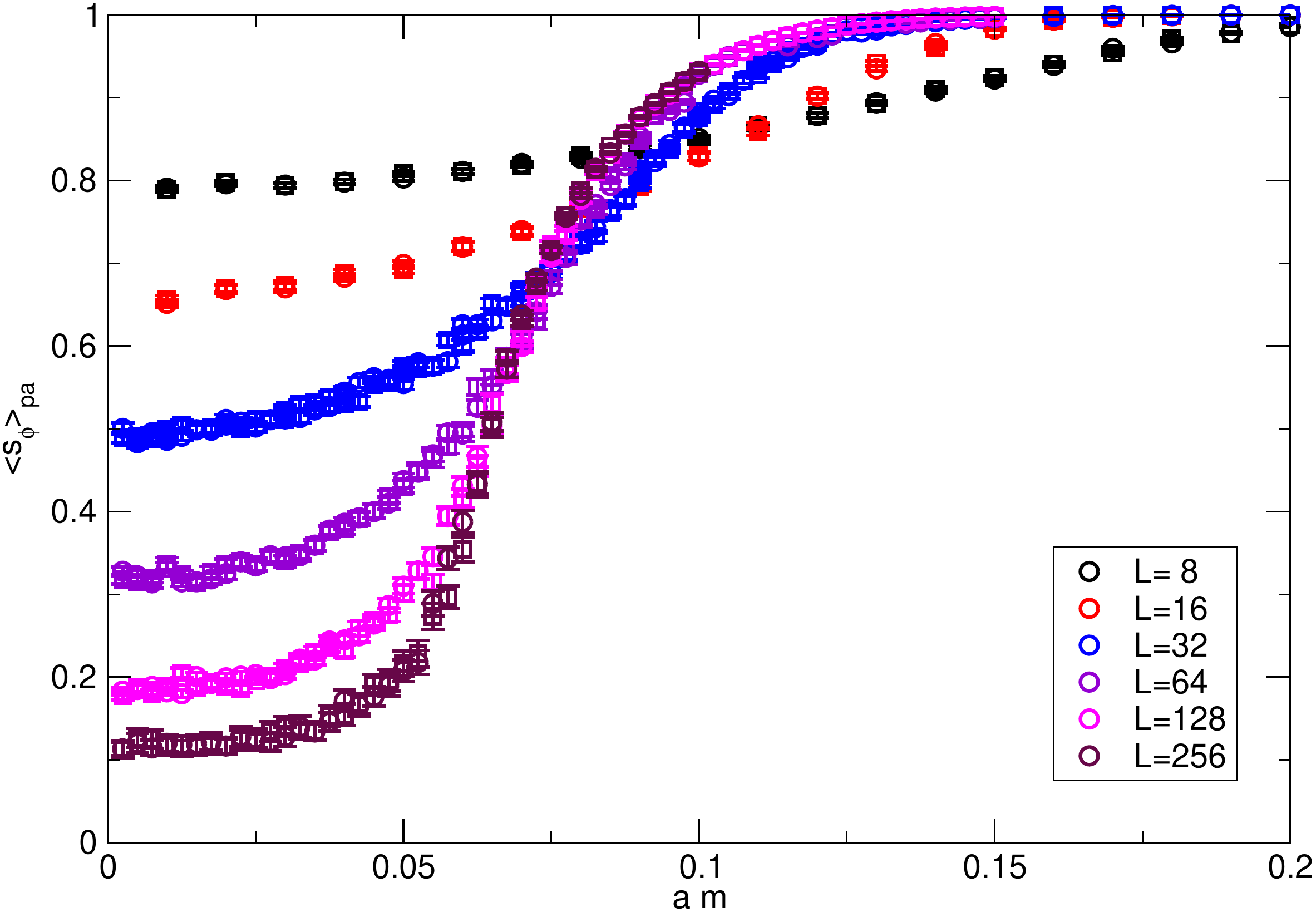} 
\caption{Ratio $Z_\text{pp}/Z$ (left plot) and $\langle s_\phi \rangle
  = \langle \text{sign} \overline \phi \rangle$ serving as a
  (pseudo-)order parameter for the supersymmetry breaking phase
  transition and the $\mathbb{Z}(2)$-symmetry breaking phase
  transition, respectively, as a function of $am$
  at fixed lattice spacing $a g = 0.03125$ for various lattice extents.  }
\label{fig:transitionscan}
\end{figure}
transition from the $\mathbb{Z}(2)$ symmetric and supersymmetry broken
phase to the $\mathbb{Z}(2)$ broken and supersymmetric phase occurs. This
is most easily done by scanning $a m$ at fixed lattice spacing $a g$
for various lattice extents $L$. The critical value $a m_c$ where the
phase transition occurs determines the dimensionless critical coupling
$m_c/g$ at the given lattice spacing. The procedure is illustrated in
Fig.~\ref{fig:transitionscan} where the left plot shows
the ratio $Z_\text{pp}/Z$ serving as a (pseudo-)order parameter for
the supersymmetry breaking phase transition, while the right plot
shows $\langle s_\phi \rangle_\text{pa} = \langle \text{sign} \overline
\phi \rangle_\text{pa}$ as a
(pseudo-)order parameter for the $\mathbb{Z}(2)$ symmetry
breaking\footnote{Note that since $Z_\text{pp} \simeq 0$ in the
  supersymmetry broken phase, expectation values need to be calculated
  in the thermal ensemble in order to be under good numerical
  control.},  as a function of the bare mass at fixed lattice spacing $a g =
0.03125$ for various lattice extents.
We note that the behaviour of $s_\phi$ seems to suggest a second order phase
transition.
Setting up the model with the
Wilson derivative for bosons and fermions yields a supersymmetric
continuum limit \cite{Golterman:1988ta}. Since the model is
 superrenomalisable it is
\begin{wrapfigure}{r}{0.5\linewidth}
\includegraphics[width=0.5\textwidth]{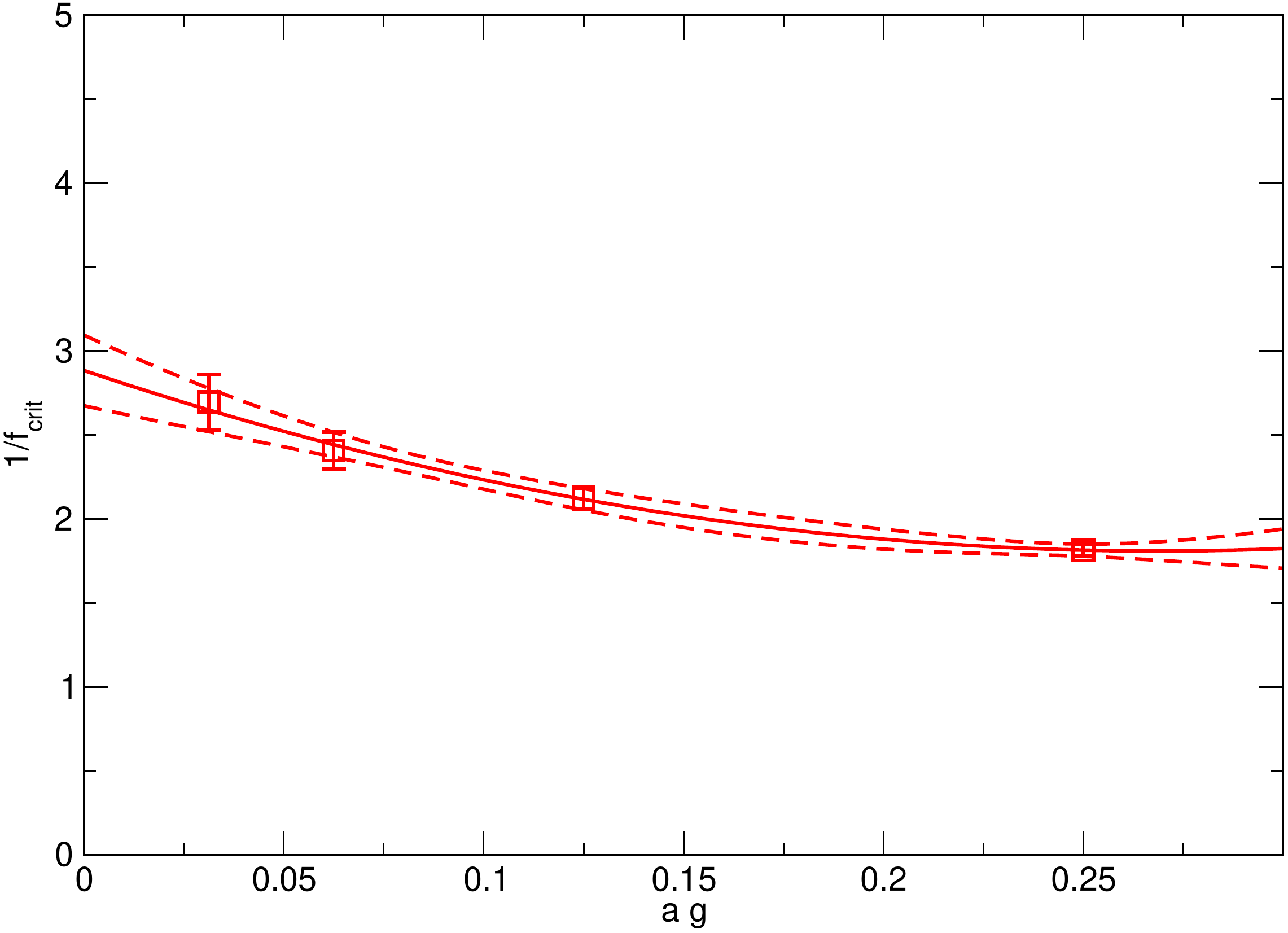} 
\caption{Continuum limit of the renormalised dimensionless critical coupling
  $f_\text{crit}$ for the supersymmetry and $\mathbb{Z}(2)$-symmetry breaking
phase transition.
}
\label{fig:mc_continuumlimit}
\end{wrapfigure} 
sufficient to tune only the mass parameter $m$ in order to obtain a
renormalised theory. Using the relation $ m^2 = m_R^2 + 2 g^2/\pi \ln m_R^2$
between the bare mass $m$ and the renormalised one $m_R$, and setting
the scale, i.e.~the lattice spacing $a$, by the dimensionful coupling
$g = \hat g/a$, one can determine the continuum limit of a
dimensionless critical coupling for the supersymmetry breaking
transition with 
\[
f_\text{crit} = \lim_{\hat g \rightarrow 0} \left. \frac{g}{m_R}\right|_\text{crit} \, .
\]
The procedure is illustrated in Fig.~\ref{fig:mc_continuumlimit} and
it will be interesting to see how this result compares to previous
determinations \cite{Wozar:2011gu,Beccaria:2004ds}.

\section{Outlook}
There are a several obvious ways to proceed from this first,
preliminary investigation. Firstly, one can use Ward identities as
(pseudo-)order parameters to determine the phase transition
point. Secondly, one can determine the boson and fermion mass
spectra. The latter is particularly simple in the fermion loop
formulation.
Thirdly, one can try to perform non-perturbative renormalisation
using the boson or fermion masses. 
Finally, it would
also be interesting to implement the loop formulation of the model
with a domain wall or overlap type fermion discretisation for which the
discrete $\mathbb{Z}(2)$ chiral symmetry remains exact at finite lattice spacing.

\bibliographystyle{JHEP}
\bibliography{sotltn1WZm}

\end{document}